\documentclass[superscriptaddress,preprintnumbers,amsmath,amssymb,prd,nofootinbib,preprint]{revtex4-1}
\pdfoutput=1
\usepackage{graphicx}
\usepackage{epstopdf}
\usepackage{dcolumn}% Align table columns on decimal point
\usepackage{bm}% bold math
\usepackage{hyperref}
\usepackage{color}
\usepackage{amsmath}
\usepackage{cancel}
\usepackage{xpatch}
\makeatletter
\def\l@subsubsection#1#2{}
\patchcmd{\@ssect@ltx}
    {\addcontentsline{toc}{#1}{\protect\numberline{}#8}}
    {}
    {}
    {}
\makeatother
\begin{document}
%%%%%%%%%%%%%%%%%%%%%%%%%%%%%%%%%%%%%%%%%%%

\def\a{\alpha}
\def\b{\beta}
\def\c{\varepsilon}
\def\d{\delta}
\def\e{\epsilon}
\def\f{\phi}
\def\g{\gamma}
\def\h{\theta}
\def\k{\kappa}
\def\l{\lambda}
\def\m{\mu}
\def\n{\nu}
\def\p{\psi}
\def\q{\partial}
\def\r{\rho}
\def\s{\sigma}
\def\t{\tau}
\def\u{\upsilon}
\def\v{\varphi}
\def\w{\omega}
\def\x{\xi}
\def\y{\eta}
\def\z{\zeta}
\def\D{\Delta}
\def\G{\Gamma}
\def\H{\Theta}
\def\L{\Lambda}
\def\F{\Phi}
\def\P{\Psi}
\def\S{\Sigma}

\def\o{\over}
\def\beq{\begin{align}}
\def\eeq{\end{align}}
\newcommand{\gsim}{ \mathop{}_{\textstyle \sim}^{\textstyle >} }
\newcommand{\lsim}{ \mathop{}_{\textstyle \sim}^{\textstyle <} }
\newcommand{\vev}[1]{ \left\langle {#1} \right\rangle }
\newcommand{\bra}[1]{ \langle {#1} | }
\newcommand{\ket}[1]{ | {#1} \rangle }
\newcommand{\EV}{ {\rm eV} }
\newcommand{\KEV}{ {\rm keV} }
\newcommand{\MEV}{ {\rm MeV} }
\newcommand{\GEV}{ {\rm GeV} }
\newcommand{\TEV}{ {\rm TeV} }
\newcommand{\1}{\mbox{1}\hspace{-0.25em}\mbox{l}}
\newcommand{\headline}[1]{\noindent{\bf #1}}
\def\diag{\mathop{\rm diag}\nolimits}
\def\Spin{\mathop{\rm Spin}}
\def\SO{\mathop{\rm SO}}
\def\O{\mathop{\rm O}}
\def\SU{\mathop{\rm SU}}
\def\U{\mathop{\rm U}}
\def\Sp{\mathop{\rm Sp}}
\def\SL{\mathop{\rm SL}}
\def\tr{\mathop{\rm tr}}
\def\mpl{M_{\rm Pl}}

\def\IJMP{Int.~J.~Mod.~Phys. }
\def\MPL{Mod.~Phys.~Lett. }
\def\NP{Nucl.~Phys. }
\def\PL{Phys.~Lett. }
\def\PR{Phys.~Rev. }
\def\PRL{Phys.~Rev.~Lett. }
\def\PTP{Prog.~Theor.~Phys. }
\def\ZP{Z.~Phys. }

\def\dd{\mathrm{d}}
\def\ff{\mathrm{f}}
\def\BH{{\rm BH}}
\def\inf{{\rm inf}}
\def\ev{{\rm evap}}
\def\eq{{\rm eq}}
\def\SM{{\rm sm}}
\def\Mpl{M_{\rm Pl}}
\def\GeV{{\rm GeV}}
\newcommand{\Red}[1]{\textcolor{red}{#1}}
\newcommand{\TL}[1]{\textcolor{blue}{\bf TL: #1}}

%%%%%%%%%%%%%%%%%%%%%%%%%%%%%%%%%%%%%%%%%%%%%%%%%%%%%%%%%%%%%%%
%\begin{titlepage}
%\begin{center}

\begin{center}
\vspace{1cm}
{\Large \bf Flavour and Higgs compositeness: \\
present and "near" future} \\
\vspace{2cm}
{\large Riccardo Barbieri} \\

{\it \small Scuola Normale Superiore, Piazza dei Cavalieri 7, 56126 Pisa, Italy and INFN, Pisa, Italy} \\
\vspace{1cm}
\end{center}
Flavour physics, from now up to the operation of the next high energy collider,  will be an important tool for BSM searches at the TeV scale. Although far from exhaustive, a particularly relevant case is represented by the possibility that the Higgs be a composite Pseudo-Nambu-Goldstone-Boson  (PNGB) at a scale $l_H=1/m_*$.
While a totally model-independent assessment of the potential of flavour physics in this case is impossible, here we  illustrate what is likely to be a minimal sensitivity on $m_*$ by considering suitable  examples.\\

%\vspace{4cm}
%\begin{center}
%A contribution to:\\
%"From my vast repertoire: the legacy of Guido Altarelli"\\
%S. Forte, A. Levy and G. Ridolfi,eds.
%\end{center}

\newpage

\section{Introduction}

Among the indirect tests of new physics, those  that aim at seeing evidence for (or setting limits on) the scale of Higgs compositeness, $l_H = 1/m_*$, as usually called in the literature, are among the most important ones, if not the dominant at all. The purpose of this note is to describe the sensitivity to $m_*$ of flavour physics with an eye to the progress foreseen before the operation of the next collider, be it an $e^+e^-$ or a $pp$ or even a muon collider, here dubbed "near" future.  A comprehensive detailed analysis of foreseen experimental and theoretical developments in flavour physics within this temporal range is definitely beyond the scope of this note. Rather we focus our attention on  an indispensable set of hypotheses  required to describe flavour in composite Higgs models, in order to see their impact on the sensitivity to $l_H = 1/m_*$ in minimal examples.

\section{Setting the framework}
\label{framework}

We assume that a new strong interaction with a confinement scale $m_*$ and a strong coupling $1< g_* <4\pi$ gives rise, after spontaneous symmetry breaking, to the Higgs, $H$, as a Pseudo-Nambu-Goldstone-Boson (PNGB). This is the scheme adopted also when discussing the ElectroWeak Precision Tests or other flavour-less precision tests in this context.  Here we are not concerned with the issue of the relation between $m_*$ and $m_H$.

The standard fermions do not feel directly this new strong interaction, but, to get a mass, they have to be connected in some (unspecified) way  with a composite operator of the strong sector, $\mathcal{O}_H$ with $<0|\mathcal{O}_H|H>\neq 0$. To describe flavour, after integrating out all states at or above a flavour scale $\Lambda_F$, the standard fermions, $f^a_i$, will enter the effective Lagrangian  below $\Lambda_F$ always multiplied by a dimensionless coupling $\lambda^a_i$. There can be more than one flavour scale $\Lambda_F$, in which case we concentrate on the scale, that we shall keep calling $\Lambda_F$, at which the top quark acquires its mass~\cite{Panico:2016ull}. (See also~\cite{KerenZur:2012fr,Barbieri:2012tu} and references therein.)
Below this scale we take as relevant effective Lagrangian 
\begin{equation}
\mathcal{L}^{top}= \frac{\Lambda_F^4}{g_{\Lambda_F}^2}[\mathcal{L}^0(\frac{\lambda^t_Lq_{L3} }{\Lambda_F^{3/2}}, \frac{\lambda^t_R t_{R}}{\Lambda_F^{3/2}},\frac{D_\mu}{\Lambda_F}, \frac{\mathcal{O}_H}{\Lambda_F^{d_H}})
+\frac{g_{\Lambda_F}^2}{16\pi^2}\mathcal{L}^1(\frac{\lambda^t_L q_{L3}}{\Lambda_F^{3/2}}, \frac{\lambda^t_R t_{R}}{\Lambda_F^{3/2}},\frac{D_\mu}{\Lambda_F}, \frac{\mathcal{O}_H}{\Lambda_F^{d_H}})+\dots]
\label{topL}
\end{equation}
as it arises from Naive Dimensional Analysis if a single new coupling $g_{\Lambda_F}$ is involved, other than $\lambda^t_{L,R}$, and the operator $\mathcal{O}_H$ has anomalous dimension $d_H$. This Lagrangian fixes in particular the effective top Yukawa interaction at $\Lambda_F$
\begin{equation}
\mathcal{L}^{top}_Y(\Lambda_F)=\frac{x_L x_R}{\Lambda_F^{d_H-1}}  \bar{q}_{L3} \mathcal{O}_H t_R,\quad\quad
x_{L,R}=\frac{\lambda^t_{L,R}}{g_{\Lambda_F}}
\end{equation}
or the top Yukawa coupling $y_t$ at the compositeness scale $m_*$, after $\mathcal{O}_H\rightarrow g_* m_*^{d_H-1} H$,
\begin{equation}
y_t=g_* x_L x_R (\frac{m_*}{\Lambda_F})^{d_H-1}.
\label{topY}
\end{equation}

As it will be used in the following, the Lagrangian (\ref{topL}) fixes as well all the effective interactions among $q_{L3}, t_R$, but not the entire set of Yukawa couplings $Y_{U,D,E}$, which, as said, will involve all the other $\lambda_i^a$ and possibly other flavour scales. As such, the full Yukawa couplings are model dependent. On the other hand the mixings of the third generation quarks, $q_{L3}$ and $t_R$ entering in (\ref{topL}) with the lighter generations, required to diagonalise these full Yukawa couplings, are crucial to compare the predictions of (\ref{topL}) in flavour experiments. To overcome this problem, calling $U_{L,R}$ and $D_{L,R}$ the unitary matrices that diagonalise $Y_U$ and $Y_D$ respectively,  we shall consider the following 3 cases:
\begin{itemize}
\item Case 1: $D_L = V, U_L =\bold{1}, D_R = U_R=\bold{1}$
\item Case 2: $U_L = V^+, D_L =\bold{1}, D_R = U_R=\bold{1}, $
%\item Case 3: $D_L =  V, U_L =\bold{1}, D_R =V,  U_R= \bold{1}$
\item Case 3: $U_L = V^+, D_L =\bold{1}, D_R =\bold{1}, U_R= V^+$
\end{itemize}
where   $V= U_L^+D_L$ is the CKM matrix.

We think that the overall consideration of these examples  illustrates the power of flavour physics. It is easy to consider motivated cases more constraining on $m_*$, whereas, on the contrary, it is hard to conceive a situation less constraining  on $m_*$  than each of these examples, especially Case 1 and 2.
  
\section{$\Delta F =2$}

From the Lagrangian (\ref{topL}), upon use of eq. (\ref{topY}), the 4-Fermi interactions at $m_*$ of $q_{L3},t_R$ among  themselves  are 
\begin{equation}
\mathcal{L}^{4F}=\frac{y_t^2}{m_*^2} (\frac{\Lambda_F}{m_*})^{2(d_H-2)}
[x_t^2 (\bar{q}_{L3}\gamma_\mu  q_{L3})^2+ (\bar{q}_{L3}t_R)(\bar{t}_R q_{L3}) +\frac{1}{x_t^2} (\bar{t}_R\gamma_\mu t_R)^2],\quad x_t = \frac{x_L}{x_R}.
\label{L4F}
\end{equation}
After going to the physical bases, this Lagrangian generates in the three cases defined above the following $\Delta F =2$ effective Lagrangians:

\begin{equation}
\mathcal{L}^{\Delta F =2}_{Case 1} = \frac{y_t^2}{m_*^2} C
[x_t^2 (\bar{d}_{Li} \xi^d_{ij}\gamma_\mu d_{Lj})^2],   \quad\quad \xi^d_{ij} = V_{tj}V^*_{ti}
\end{equation}

\begin{equation}
\mathcal{L}^{\Delta F =2}_{Case 2} = \frac{y_t^2}{m_*^2} C
[x_t^2 (\bar{u}_{Li} \xi^u_{ij}\gamma_\mu u_{Lj})^2],   \quad\quad \xi^u_{ij} = V_{ib}V^*_{jb}
\end{equation}

\begin{equation}
\mathcal{L}^{\Delta F =2}_{Case 3} = \frac{y_t^2}{m_*^2} C
[x_t^2 (\bar{u}_{Li} \xi^u_{ij}\gamma_\mu u_{Lj})^2 + (\bar{u}_{Li} \xi^u_{ij} u_{Rj})(\bar{u}_{Rk} (\xi^u_{lk})^* u_{Ll})+
\frac{1}{x_t^2} (\bar{u}_{Ri} \xi^u_{ij}\gamma_\mu u_{Rj})^2]
\end{equation}
The overall coefficient 
\begin{equation}
C = (\frac{\Lambda_F}{m_*})^{2(d_H-2)}
\end{equation}
is bigger than one for any scale $\Lambda_F> m_*$ since $d_H \geq 2$~\cite{Rattazzi:2008pe}.

%From these effective Lagrangian a full fit of current flavour data~\cite{Silvestrini:2019sey}  allows to set the bounds on the compositeness scale $m_*$ summarised in Table~\ref{DeltaF=2}. In the "near" future these bounds are expected to improve by about a factor of 2 in the first line, limited by the parametric uncertainty on the CKM angles as determined by tree level measurements, and by a factor of about 3 in $\Delta C=2$, i.e. on the second and third lines.
From these effective Lagrangians a full fit of current flavour data~\cite{Silvestrini:2019sey}  allows to set the bounds on the compositeness scale $m_*$ summarised in Table~\ref{DeltaF=2}. In the "near" future these bounds are expected to go to the values indicated in parenthesis.
 \begin{table}[t]
$$\begin{array}{c|c|c|c|c}
&\Delta S=2&\Delta C=2&\Delta B_d=2&\Delta B_s=2\\ \hline
Case 1&7(13)x_t&-&8(20)x_t&9(20)x_t\\ \hline
Case 2&-&3(10)x_t&-&-\\ \hline
Case 3&-&10(30)&-&-\\ \hline
\end{array}$$
\caption{Summary of the $95\%$ probability lower bounds  on $m_*/TeV$ from $\Delta F=2$. $x_t$ can vary in the  range  $y_t/g_{\Lambda_F} < x_t <g_{\Lambda_F}/y_t$. In parenthesis the sensitivity expected at the end of LHC is indicated in the different cases. }
\label{DeltaF=2}
\end{table}

\section{$\Delta  F=1$}

By analogous considerations to the ones developed in the previous Section, the effective Lagrangian most relevant to $\Delta  F=1$ transitions is
\begin{equation}
\mathcal{L}^{\Delta F=1}=\frac{g_*y_t}{m_*^2}C^{1/2}i (H^\dag \overleftrightarrow{D_\mu} H)  [x_t(\bar q_{L3} \gamma^\mu q_{L3})+\frac{1}{x_t}(\bar t_{R} \gamma^\mu t_{R})]
\end{equation}

\begin{figure}[t]
\centering
\includegraphics[clip,width=.48\textwidth]{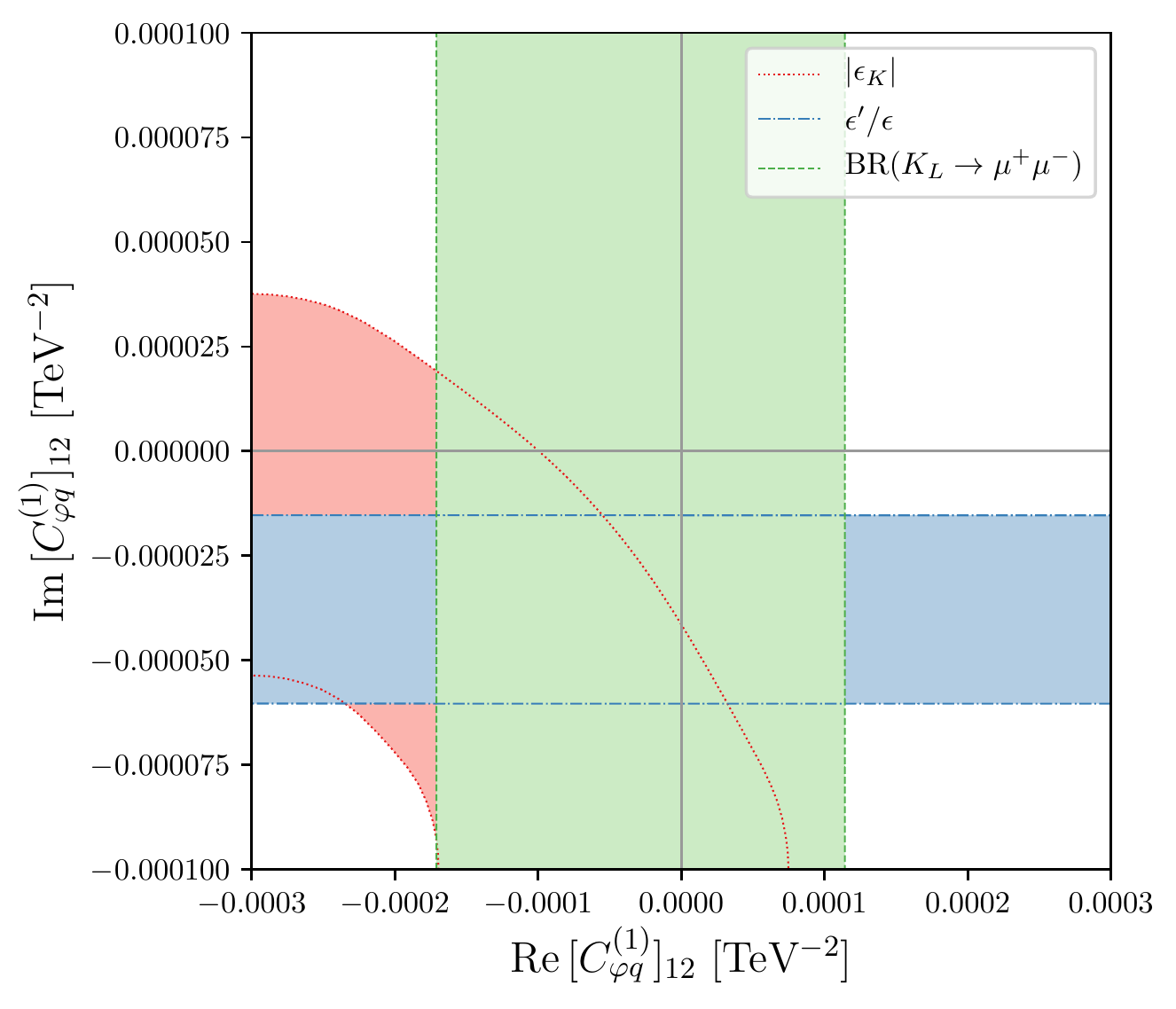}
\includegraphics[clip,width=.66\textwidth]{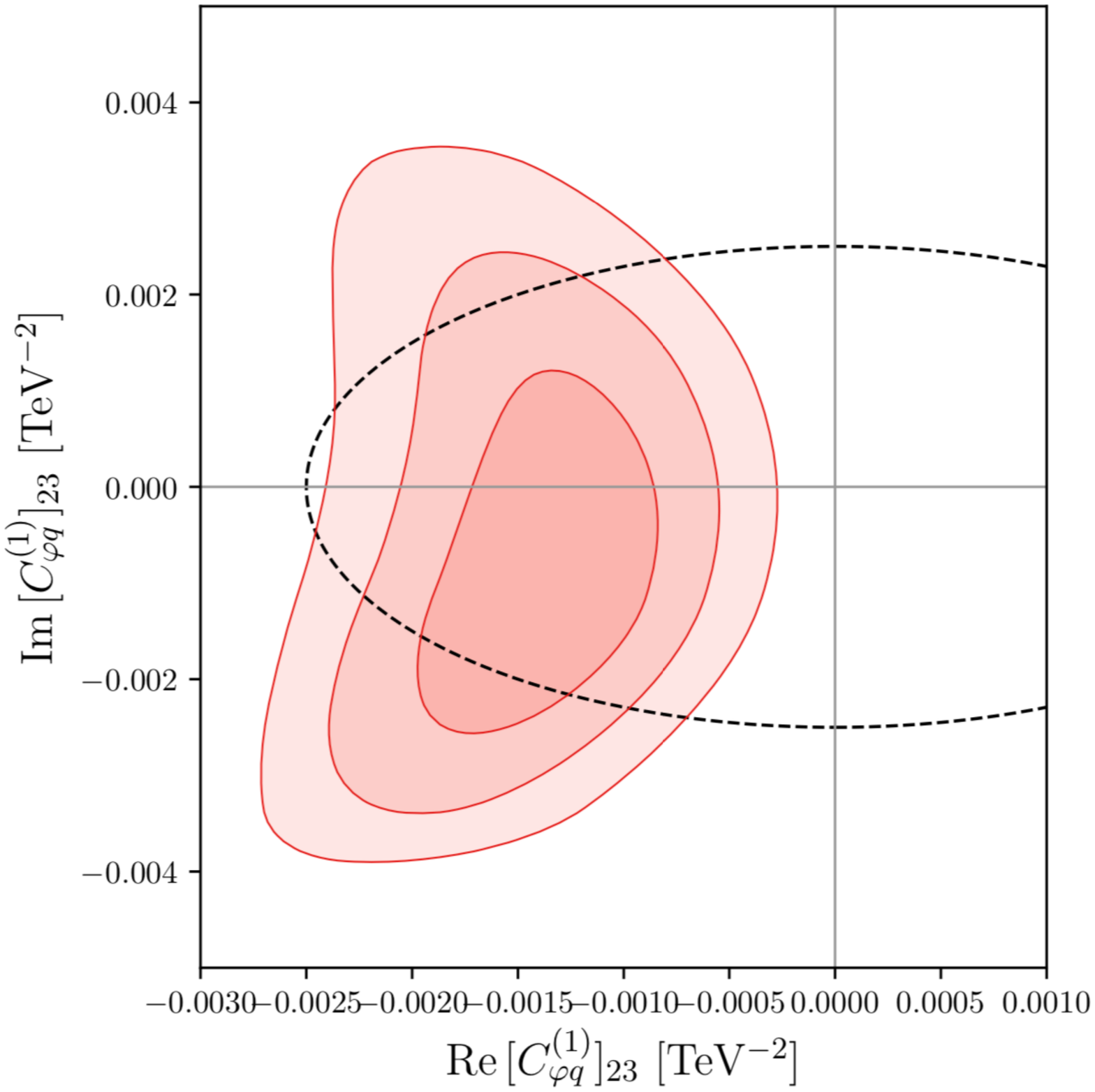}
\caption{Current constraints on the coefficient $C^{(1)}_{\phi q}|_{12}$, upper figure, and on the coefficient $C^{(1)}_{\phi q}|_{23}$, lower figure. See text. Courtesy of David Straub.
}
\label{fig:DeltaF=1}
\end{figure}

From this Lagrangian the most relevant bounds are obtained for Case 1, i.e, after going to the physical basis, from
\begin{equation}
\mathcal{L}^{\Delta F=1}_{Case 1}=\frac{g_*y_t}{m_*^2}C^{1/2}i (H^\dag \overleftrightarrow{D_\mu} H)  [x_t(\bar d_{Li} \xi_{ij}^d\gamma^\mu d_{Lj})].
\label{eq:DeltaF=1}
\end{equation}
In Fig.s~\ref{fig:DeltaF=1}, we show the current constraints on the overall coefficients $C^{(1)}_{\phi q}|_{ij}$ of the operator 
$i (H^\dag \overleftrightarrow{D_\mu} H)  (\bar d_{Li} \gamma^\mu d_{Lj})$
in the $12=ds$ (upper figure) and $23=sb$ channels (lower figure). 
 The uncertainties of the observables in the $12=ds$ case are dominated by theory. An important input would come from a determination of the branching ratio of $K^+\rightarrow \pi^+ \nu\bar \nu$ at the $10\%$ level. Achieving this sensitivity would be equivalent to $Re [C^{(1)}_{\phi q}]_{12}\lesssim 4\cdot 10^{-5} TeV^{-2}$ (and might perhaps support or dilute the putative evidence for BSM contribution in $\epsilon^\prime/\epsilon$, hinted in Fig.~\ref{fig:DeltaF=1}, upper). Due to the large number of observables relevant to the $23=sb$ case, the lower figure
shows the result of an overall fit. Particularly important are the branching ratio of $B_s\rightarrow\mu^+\mu^-$ and the angular variables in $B\rightarrow K^*\mu\mu$, both statistically dominated at present~\cite{Aebischer:2018iyb}.

Based on Fig.s~\ref{fig:DeltaF=1} we consider
\begin{equation}
Im [C^{(1)}_{\phi q}]_{12}\lesssim 6\cdot 10^{-5} TeV^{-2},\quad\quad
Re [C^{(1)}_{\phi q}]_{23}, Im  [C^{(1)}_{\phi q}]_{23}\lesssim 2\cdot 10^{-3} TeV^{-2}
\end{equation}
which correspond to the bounds on $m_*$ shown in Table~\ref{DeltaF=1}.

Note that eq.~\ref{eq:DeltaF=1} contains as well a correction to the $Zb\bar b$ coupling
\begin{equation}
\delta g_{bL} = C^{(1)}_{\phi q}|_{33}\frac{v^2}{2}
\end{equation}
so that, requiring $\delta g_{bL} < 1.5\cdot 10^{-3}$, one gets $m_*\gtrsim 4.5 \sqrt{g_*x_t}~TeV$.

 \begin{table}[t]
$$\begin{array}{c|c|c}
&12&23\\ \hline
Case 1&1.7 \sqrt{g_*x_t}&4.5\sqrt{g_*x_t}\\ \hline
\end{array}$$
\caption{ Lower bounds  on $m_*/TeV$ from $\Delta F=1$ transitions. $x_t$ can vary in the range  $y_t/g_{\Lambda_F} < x_t <g_{\Lambda_F}/y_t$}
\label{DeltaF=1}
\end{table}

We do not consider  the bounds/sensitivity from $\Delta C=1$ operators due to the difficulty of estimating the SM contributions to these processes.

In some models with a suitable custodial parity~\cite{Agashe:2006at} the relation 
\begin{equation}
C^{(1)}_{\phi q}|_{33} = - C^{(3)}_{\phi q}|_{33}
\end{equation}
with the coefficient $C^{(3)}_{\phi q}|_{33}$ of the operator $i (H^\dag \overleftrightarrow{D_\mu} \sigma_aH)  (\bar d_{L3} \gamma^\mu \sigma^ad_{L3})$ may suppress all these $\Delta F=1$ effects.

\section{Dipole operators}
\subsection{Neutron Electric Dipole Moment}

From the Lagrangian (\ref{topL}) at one loop one obtains the dipole operators (identifying $\Lambda_F$ with $m_*$ and $g_{\Lambda_F}$ with $g_*$ for ease of exposition, but without influence on the bounds on $m_*$)
\begin{equation}
\mathcal{L}^{dip} = \frac{g_*^2}{16\pi^2} \frac{m_t}{m_*^2}
[C^{dip}(\bar{t}_L\sigma_{\mu\nu}t_R)e F^{\mu\nu} + \tilde{C}^{dip}(\bar{t}_L\sigma_{\mu\nu}T^at_R)g_S G^{\mu\nu}_a],
\label{dip_top}
\end{equation}
where $C^{dip}, \tilde{C}^{dip}$ are coefficients of order unity, in general complex.
In Case 3, after going in the physical basis, this gives
\begin{equation}
\mathcal{L}^{dip}_{Case 3} = \frac{g_*^2}{16\pi^2} \frac{m_t}{m_*^2}
[C^{dip}(\bar{u}_{Li}\xi_{ij}^u\sigma_{\mu\nu}u_{Rj})e F^{\mu\nu} + \tilde{C}^{dip}(\bar{u}_{Li}\xi_{ij}^u\sigma_{\mu\nu}T^a u_{Rj})g_S G^{\mu\nu}_a]
\end{equation}

The cromo-electric dipole moments of the quarks feed into the neutron  electric dipole moment, currently bound by~\cite{Baker:2006ts}
\begin{equation}
d_n < 2.9\cdot 10^{-26}~e~cm,
\end{equation}
via their contribution to the Weinberg operator.
As a consequence, from the up, charm and top contributions one gets~\cite{Sala:2013osa}
\begin{equation}
 Im(\tilde{C}^{dip})\xi^u_{11} < 1.3\cdot 10^{-8},\quad\quad Im(\tilde{C}^{dip})\xi^u_{22} <1.8\cdot 10^{-5},\quad\quad Im(\tilde{C}^{dip})\xi^u_{33} <3.3\cdot 10^{-2},
\end{equation}
with a common factor $(g_*^2/16\pi^2)(TeV/m_*)^2$ left understood  on the left  side of each of these bounds. Given the values of $\xi^u_{ii}$, the bound in Case 3 is dominated by the contribution from the up quark.
Taking $Im(\tilde{C}^{dip})=1$, these bounds translate into the bounds on $m_*$ shown in Table~\ref{EDMs}. The sensitivity to the neutron  electric dipole moment is expected to be improved by one order of magnitude in the near future by a dedicated experiment at PSI. Consequently, in absence of a signal, the limits shown in Table~\ref{EDMs} are expected to improve by about a factor of  3.

 \begin{table}[t]
$$\begin{array}{c|c}
&m_*/TeV\\ \hline
Case 1&5.5 \frac{g_*}{4\pi}\\ \hline
Case 2&5.5 \frac{g_*}{4\pi}\\ \hline
Case 3&32 \frac{g_*}{4\pi}\\ \hline
\end{array}$$
\caption{Lower bounds on $m_*/TeV$ from the neutron dipole moment.}
\label{EDMs}
\end{table}

\subsection{Electron Electric Dipole Moment}

Within the restricted framework considered so far and specified in Section~\ref{framework} the most significant effect on the electron EDM arises at two loops via the Barr-Zee-type diagrams. This proceeds through a one loop contribution to the $H^2F\tilde{F}$ operator, which is then transferred to  the electron EDM by running to the low energy scale~\cite{Panico:2018hal}. An estimate of the overall effect to the EDM $d_e$ is 
\begin{equation}
\frac{d_e}{e} \approx \frac{g_*^2}{16\pi^2}\frac{e^2}{16\pi^2}\frac{y_tx_t}{g_*}\frac{m_e}{m_*^2}
\end{equation}
where we have assumed an order one phase appearing in the coefficient of the $H^2F\tilde{F}$ operator. Requiring that this estimate be less than the recently reported result by the ACME collaboration, $|d_e|< 1.1\cdot 10^{-29}~e\cdot cm$~\cite{ACME}, leads to the bound
\begin{equation}
m_*< 6\sqrt{g_* x_t} TeV.
\end{equation}
The large effective electromagnetic fields $(> 10~ GV/cm)$ present in heavy polar molecules may allow in the coming years  a greatly improved sensitivity on the electron EDM, potentially improving  the  ACME result by orders of magnitude.

\section{Leptonic flavour}
\label{Leptons}
In all considerations developed so far no assumption had to be made about the scales at which the elementary fermions interact with  the composite Higgs sector to get their Yukawa couplings.  To discuss flavour in the case of leptons we assume that the $\tau$ lepton gets its Yukawa coupling at the same scale $\Lambda_F$ at which the top quark gets its own. In practice this means that $\mathcal{L}^0$ and $\mathcal{L}^1$ in eq. (\ref{topL}) acquire a further dependence on $
\lambda^\tau_L l_{L3}/\Lambda_F^{3/2}, \lambda^\tau_R \tau_R/\Lambda_F^{3/2}$. 
Furthermore we assume that the mixing matrices in the leptons are related to their masses $(i,j=e,\mu, \tau)$ by
\begin{equation}
|V^{Ll}_{i>j}| = |V^{Rl}_{i>j}| = \frac{m_j}{m_i}
\end{equation}

In full analogy with the case of the quarks, in particular eq. (\ref{dip_top}), the operator that leads to the
strongest contraints in the lepton case is
\begin{equation}
\mathcal{L}^{dip}_\tau = \frac{g_*^2}{16\pi^2} \frac{m_\tau}{m_*^2}
C^{dip}_\tau(\bar{\tau}_L\sigma_{\mu\nu}\tau_R)e F^{\mu\nu}
\label{Ctau}
\end{equation}
i.e., after going to the physical basis,
\begin{equation}
\mathcal{L}^{dip}_l = \frac{g_*^2}{16\pi^2} \frac{m_\tau}{m_*^2}
C^{dip}_\tau[(\bar{\tau}_L\sigma_{\mu\nu}\xi^l_{\tau\mu}\mu_R)
+ (\bar{\mu}_L\sigma_{\mu\nu}\xi^l_{\tau\mu}e_R)]e F^{\mu\nu},\quad\quad \xi^l_{ij}=V^l_{\tau j}V_{\tau i}^{l*}
\end{equation}
plus similar terms with the role of L and R reversed. In view of the current bounds, this leads to the lower limits on $m_*$ shown in Table~\ref{taumugamma}. Since the sensitivity to the relevant branching ratios is expected to improve by one order of magnitude, these bounds are expected to improve by about a factor of three.
\begin{figure}[t]
\centering
\includegraphics[clip,width=.70\textwidth]{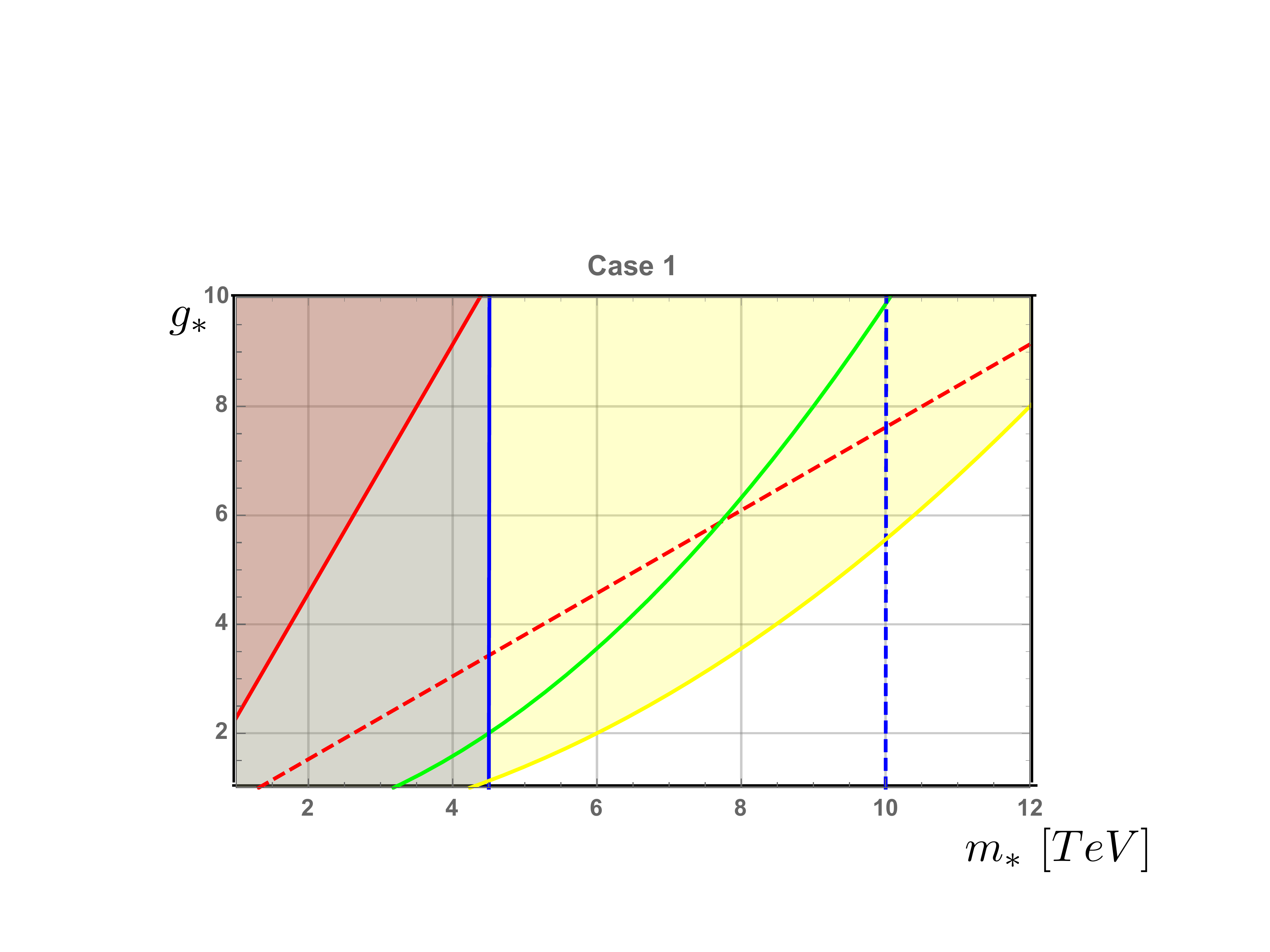}
\caption{Case 1: Current bounds  now (full lines) and the sensitivity expected in the "near" future (dotted lines)  from $\Delta B_s=2$ (blue) and the neutron EDM (red). Also shown is the current bound from the electron EDM (yellow) and from $\Delta B_s=1$  (green) in models without a custodial parity. Everywhere $x_t=1/2$, eq.~\ref{L4F}.
}
\label{fig:Case1}
\end{figure}

 \begin{table}[t]
$$\begin{array}{c|c|c}
&\tau\rightarrow \mu \gamma&\mu\rightarrow e \gamma \\ \hline
m_*/TeV&8\frac{g_*}{4\pi}&16\frac{g_*}{4\pi}\\ \hline
\end{array}$$
\caption{Lower bounds on $m_*/TeV$ from LFV decays}
\label{taumugamma}
\end{table}

If $C^{dip}$ has a phase, again after going to the physical basis, eq. (\ref{Ctau}) leads to electric dipole moments for the leptons. For the electron,
\begin{equation}
\mathcal{L}^{dip}_e = \frac{g_*^2}{16\pi^2} \frac{m_\tau}{m_*^2}
C^{dip}_\tau (\frac{m_e}{m_\tau})^2 (\bar{\tau}_L\sigma_{\mu\nu}\tau_R)e F^{\mu\nu},
\label{Ce}
\end{equation}
so that, with maximal phase, the current limit on the $eEDM < 1.1\cdot 10^{-29}~e\cdot cm$, leads to the bound
\begin{equation}
m_* < 20\frac{g_*}{4\pi} TeV.
\end{equation}

\section{Summary}

\begin{figure}[t]
\centering
\includegraphics[clip,width=.70\textwidth]{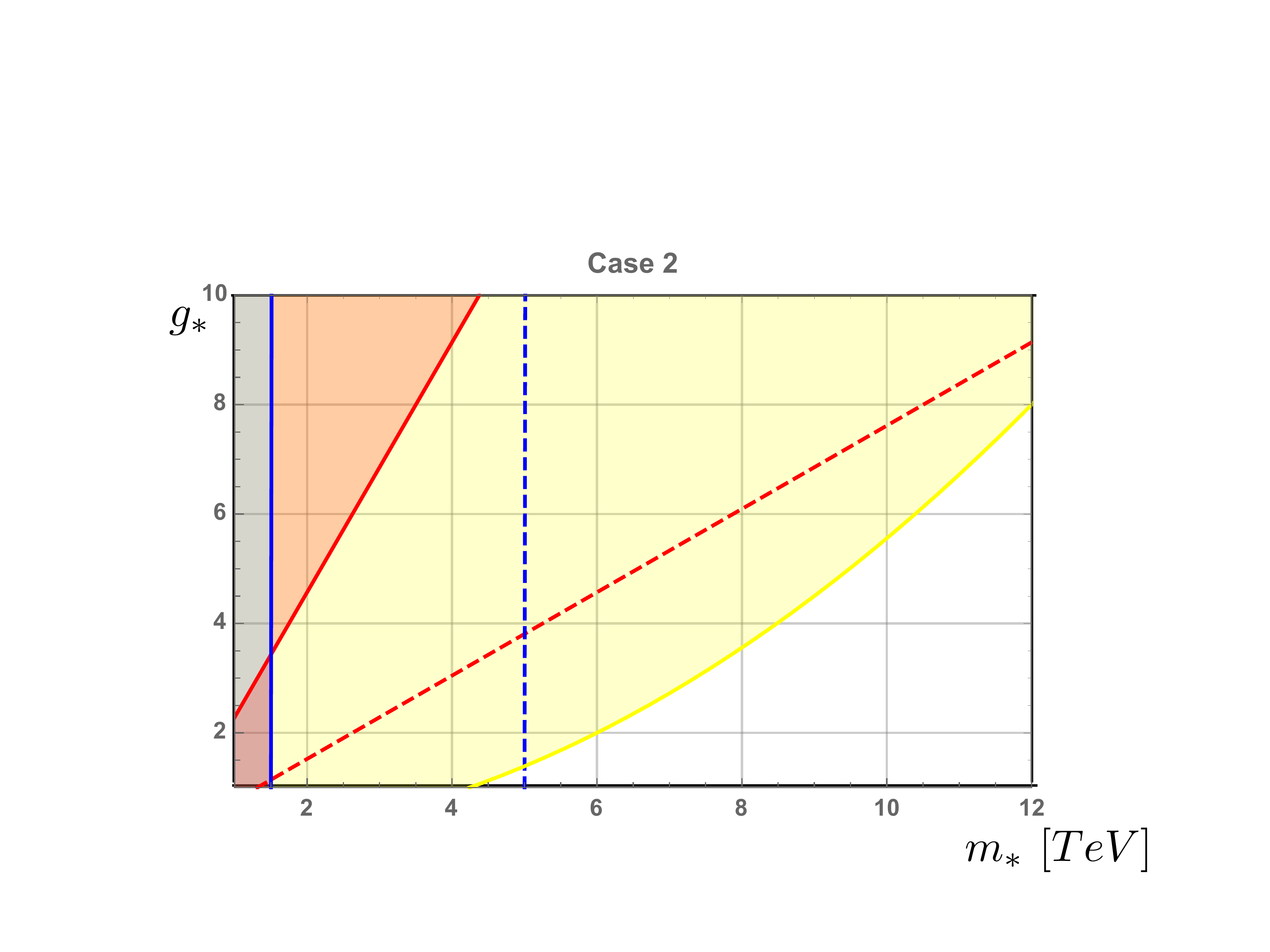}
\caption{Case 2: Current bounds  now (full lines) and the sensitivity expected in the "near" future  (dotted lines)  from $\Delta C=2$ (blue) and the neutron EDM (red). Also shown is the current bound from the electron EDM (yellow). Everywhere $x_t=1/2$, eq.~\ref{L4F}.}
\label{fig:Case2}
\end{figure}

\begin{figure}[t]
\centering
\includegraphics[clip,width=.70\textwidth]{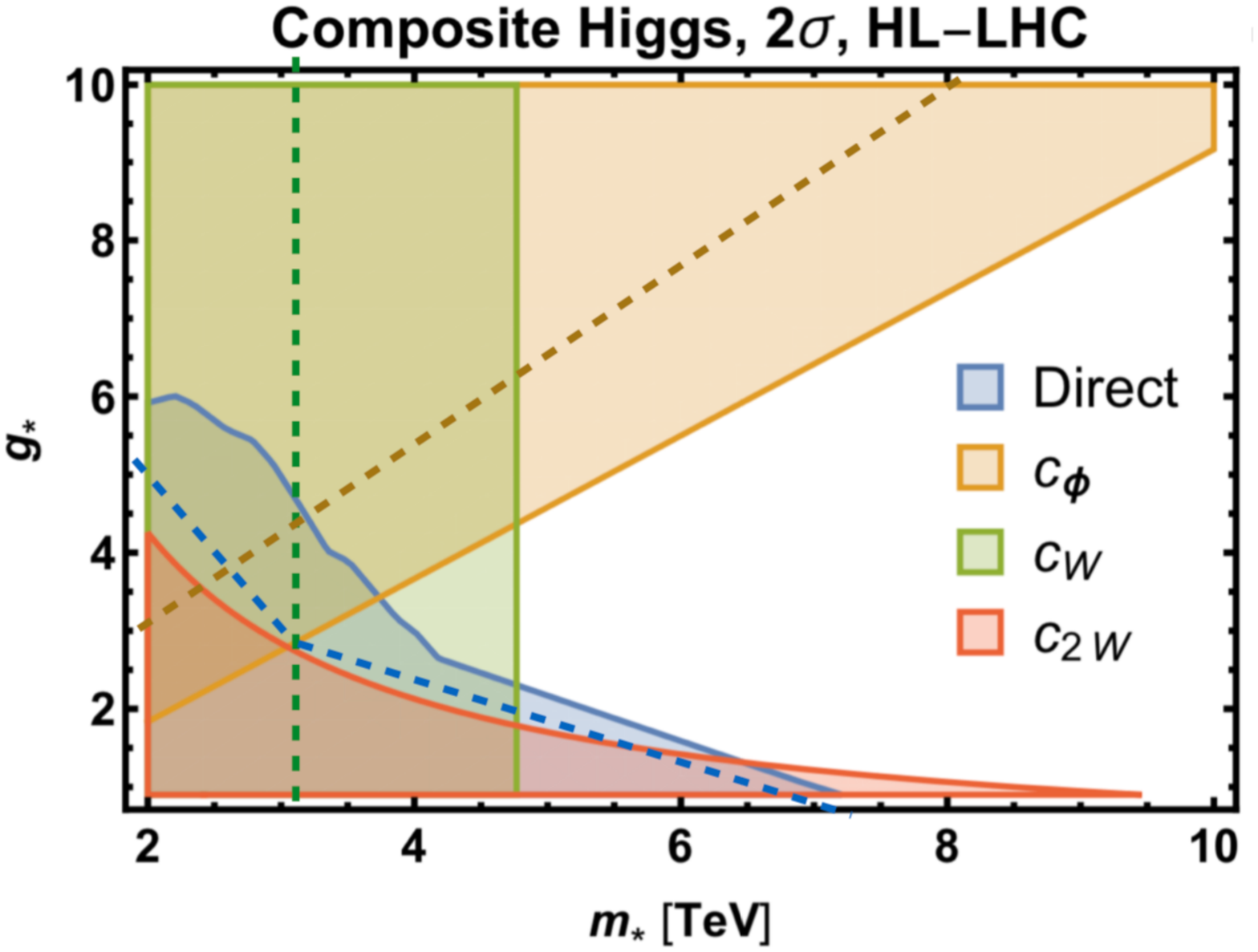}
\caption{ Current bounds  now (dotted lines) and the sensitivity expected at the end of HL-LHC (full lines) from flavour-less  Precision Tests and direct searches. Adapted from a talk by  A. Wulzer for the European Strategy, Granada,13-16 May, 2019}
\label{fig:EWPT}
\end{figure}

From now up to the operation of the next high energy accelerator, flavour physics will be an important tool for BSM searches at the TeV scale. A particularly relevant case is represented by the possibility that the Higgs be a composite PNGB at a scale $l_H=1/m_*$.
While a totally model-independent assessment of the potential of flavour physics  is impossible, one can nevertheless consider two  examples, Case 1 and 2 defined in Section~\ref{framework}, that illustrate what is likely to be a minimal sensitivity on $m_*$. This is summarised in Fig.s~\ref{fig:Case1} and~\ref{fig:Case2} for the two cases respectively. Other cases considered in the text (Case 3 in Section~\ref{framework}, and Section~\ref{Leptons}) give stronger and/or additional constraints. For comparison we show in Fig.~\ref {fig:EWPT} the sensitivity expected from flavour-less  Precision Tests on the same basic composite Higgs model. Taking into account that in all the three figures the various bounds can be moved by different $\mathcal{O}(1)$ factors, the complementarity of the two approaches is manifest.

Although  the considerations developed in this note are far from exhausting the potential impact of flavour physics in the "near" future, with the general aim of finding clues to attack the flavour puzzle, we think that they illustrate concretely such potential in a particularly relevant example\footnote{One may wonder if and how the putative anomalies  currently observed in B-decays can fit into a composite Higgs picture. They could potentially fit in Case 2 with  $m_*\gtrsim 2$ TeV and $g_*=(1\div 2)m_*/TeV$, Fig.\ref{fig:Case2} (as well as $x_\tau = \lambda^\tau_L/\lambda^\tau_R$ close to $ g_*/y_\tau$) with a suppressed CP-violating phase in the electron EDM. 
}.

%
%\appendix
%\section{Full Right compositeness}

\begin{acknowledgments}
I am grateful to Gino Isidori, Luca Silvestrini and in particular David Straub for their helpful comments and informations.

\end{acknowledgments}

%%%%%%%%%%5\appendix

\begingroup
\renewcommand{\addcontentsline}[3]{}% Remove functionality of \addcontentsline
\renewcommand{\section}[2]{}% Remove functionality of \section

\endgroup
  
\end{document}